\documentclass{article}

\PassOptionsToPackage{numbers, compress}{natbib}

\usepackage[final]{neurips_2024}
\usepackage[pdftex]{graphicx}

\usepackage[utf8]{inputenc} 
\usepackage[T1]{fontenc}    
\usepackage{url}            
\usepackage{booktabs}       
\usepackage{amsfonts}       
\usepackage{nicefrac}       
\usepackage{microtype}      
\usepackage{xcolor}         
\usepackage{booktabs}
\usepackage{graphicx}
\usepackage{booktabs}
\usepackage{multirow}
\usepackage{caption}
\usepackage{subcaption} 
\usepackage{graphicx}
\usepackage{booktabs}
\usepackage{multirow}
\usepackage{makecell}
\usepackage{amsmath}
\usepackage{threeparttable}
\usepackage{enumitem}

\setlist[itemize]{leftmargin=*}
\setlist[enumerate]{leftmargin=*}
\usepackage{xtab}

\usepackage[accsupp]{axessibility}  
\definecolor{citeblue}{RGB}{48,111,186}
\usepackage[pagebackref=false,breaklinks=true,colorlinks=true,citecolor=citeblue,bookmarks=false]{hyperref}
\usepackage{multirow}
\usepackage{colortbl}
\usepackage{cleveref}

\crefname{figure}{Fig.}{Figs.}
\crefname{table}{Tab.}{Tabs.}
\crefname{equation}{Eqn.}{Eqns.}

\definecolor{darkred}{rgb}{0.7, 0.0, 0.0}
\definecolor{darkred2}{rgb}{0.5, 0.0, 0.0}
\definecolor{darkred3}{rgb}{0.9, 0.0, 0.0}
\definecolor{darkgreen}{rgb}{0.0, 0.42, 0.24}
\definecolor{darkblue}{rgb}{0.10, 0.17, 0.8}
\definecolor{Gray}{gray}{0.93}

\title{PhoCoLens: Photorealistic and Consistent Reconstruction in Lensless Imaging}

\author{%
    Xin Cai$^{1,2}$, Zhiyuan You$^{1}$, Hailong Zhang$^3$,  Wentao Liu$^{2,4}$, Jinwei Gu$^{1}$, Tianfan Xue$^{1}$ \\
  $^1$The Chinese University of Hong Kong, $^2$Shanghai AI Laboratory, $^3$Tsinghua University, $^4$SenseTime\\
  {\tt\small \{cx023, yz023, tfxue\}@ie.cuhk.edu.hk, jwgu@cuhk.edu.hk,} \\ 
  {\tt\small \{zhanghl21\}@mails.tsinghua.edu.cn, liuwentao@sensetime.com}
}

\begin{document}

\maketitle
\vspace{-1em}
\begin{abstract}
Lensless cameras offer significant advantages in size, weight, and cost compared to traditional lens-based systems. Without a focusing lens, lensless cameras rely on computational algorithms to recover the scenes
from multiplexed measurements. However, current algorithms struggle with inaccurate forward imaging models and insufficient priors to reconstruct high-quality images. 
To overcome these limitations, we introduce a novel two-stage approach for consistent and photorealistic lensless image reconstruction. The first stage of our approach ensures data consistency by focusing on accurately reconstructing the low-frequency content with a spatially varying deconvolution method that adjusts to changes in the Point Spread Function (PSF) across the camera's field of view. The second stage enhances photorealism by incorporating a generative prior from pre-trained diffusion models. By conditioning on the low-frequency content retrieved in the first stage, the diffusion model effectively reconstructs the high-frequency details that are typically lost in the lensless imaging process, while also maintaining image fidelity. Our method achieves a superior balance between data fidelity and visual quality compared to existing methods, as demonstrated with two popular lensless systems, PhlatCam and DiffuserCam. Project website: \url{phocolens.github.io}.

\end{abstract}

\begin{figure}[h]
\centering
\includegraphics[width=\linewidth]{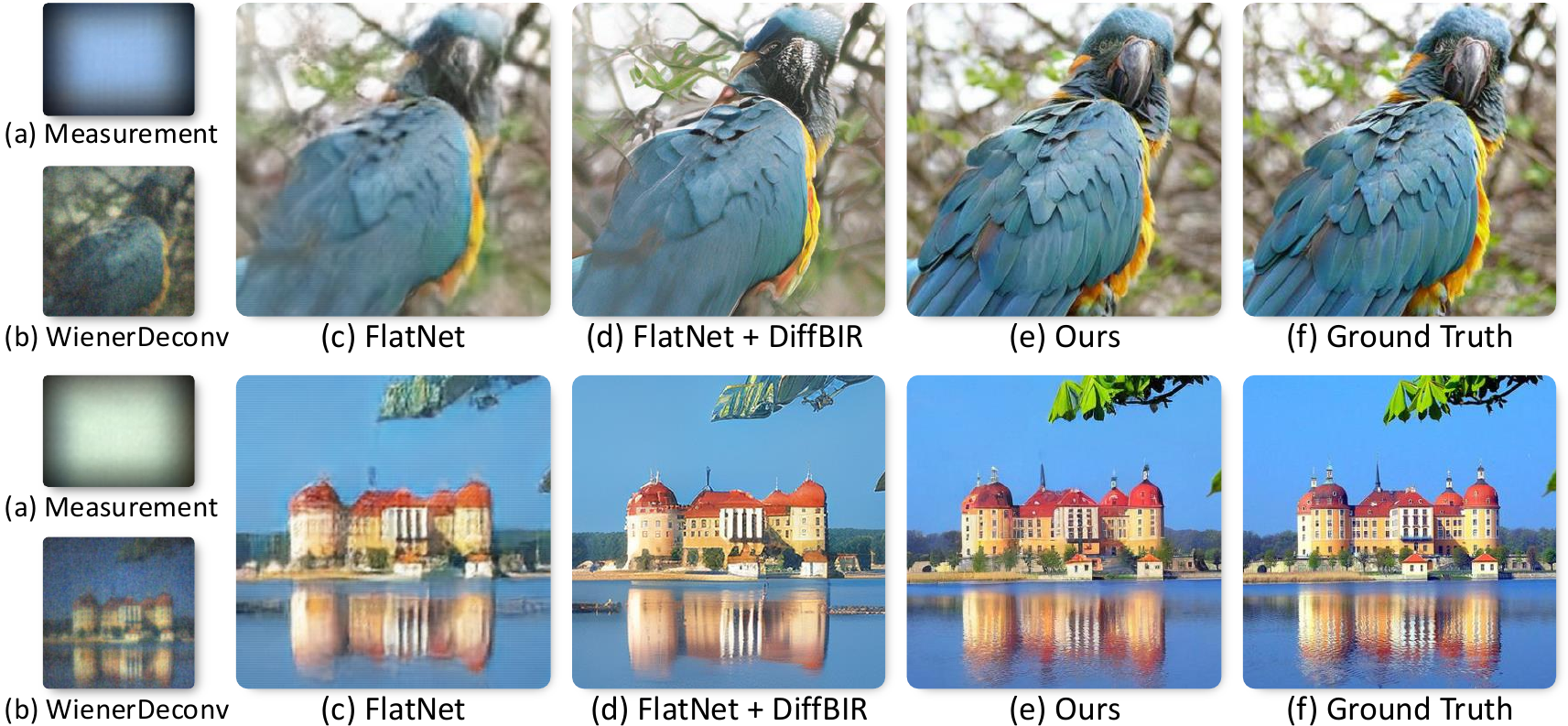}
\caption{
        We introduce PhoCoLens, a lensless reconstruction algorithm that achieves both better visual quality and consistency to the ground truth than existing methods. Our method recovers more details compared to traditional reconstruction algorithms (b) and (c), and also maintains better fidelity to the ground truth compared to the generative approach (d). 
}
\label{fig:teaser}
\end{figure}

\section{Introduction}
\vspace{-5pt}

Lensless imaging systems have emerged as a groundbreaking solution for ultra-compact, lightweight, and cost-effective imaging. These imaging systems replace lenses by amplitude \cite{asif2016flatcam,bezzam2022privacy} or phase masks \cite{antipa2018diffusercam,boominathan2020phlatcam,kuo2020chip} placed close to the sensor to modulate incoming light.
This design significantly reduces camera size and weight and enables innovative sensor shapes such as spherical \cite{hua2023angle} or cylindrical.
\begin{figure}[t]
\vspace{-5pt}
\begin{minipage}[thbp]{0.6\textwidth}
\centering
\includegraphics[width=\linewidth]{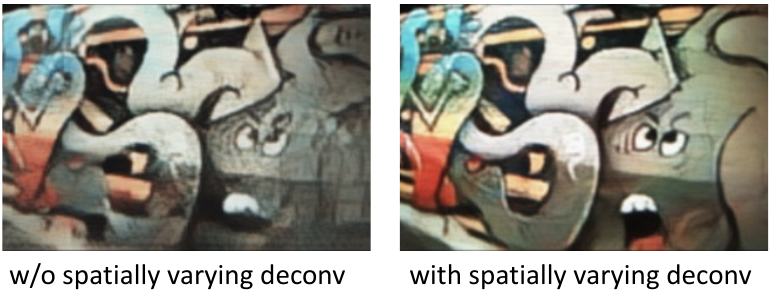}
\vspace{-15pt}
\captionof{figure}{Comparison between results without (left) and with (right) our spatially varying deconvolution.}
\label{fig:svd_comparson}
\end{minipage}
\hfill
\begin{minipage}[thbp]{0.38\textwidth}
\centering
\includegraphics[width=\linewidth]{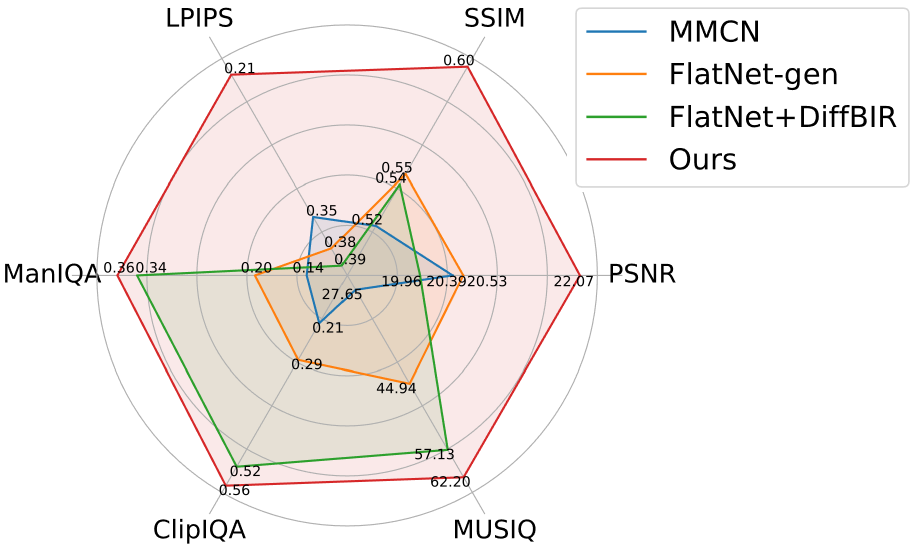}
\vspace{-15pt}
\captionof{figure}{Evaluation of fidelity and authenticity on PhlatCam~\cite{khan2020flatnet} dataset.}
\label{fig:radar}
\end{minipage}
\vspace{-15pt}
\end{figure}

Considering the raw measurements from lensless cameras are typically blurry and unrecognizable, it is hard to recover a high-quality image of the original scene.
An example of lensless measurement and recovered images by WienerDeconv \cite{wiener1949extrapolation} is shown in the left column of \cref{fig:teaser}. Due to the lack of focusing elements, lensless cameras cannot directly record the scene but encode it into a complex diffraction pattern. This encoding process can be approximated as a convolution with a large Point Spread Function (PSF). The PSF acts like a low-pass filter applied to the scene thus introducing ambiguity, which means there could be multiple possible recoveries for a single measurement. Therefore, powerful algorithms are critical for high-quality reconstruction in lensless systems.

The primary challenge of a lensless reconstruction is to achieve both photorealism and consistency. Photorealism requires high-quality reconstruction with rich details, while consistency further requires reconstructed contents to be consistent with the original scene. Traditional techniques, like WienerDeconv, can reconstruct images that align with the ground truth but visual quality is significantly degraded (\cref{fig:teaser}b). 
Learning-based approaches such as FlatNet-gen \cite{khan2020flatnet} attempt to enhance visual quality by training with paired images and lensless measurements, yet they often fail to recover high-frequency details (\cref{fig:teaser}c).
The visual quality can be further improved by the restoration algorithms using generative priors, like DiffBIR~\cite{lin2023diffbir} (\cref{fig:teaser}d). While these restoration methods can inject rich details, they may also alter image content or insert non-existent objects, breaking consistency. For example, in \cref{fig:teaser}d, the beak in the top row has the wrong shape compared to the ground truth (\cref{fig:teaser}f), and the leaves in the bottom row look fake.

Moreover, an inaccurate imaging process simulation may also hurt data consistency. Most existing reconstruction algorithms \cite{khan2020flatnet,monakhova2019learned} simplify the imaging process as a convolution with a shift-invariant PSF. However, in practice, PSFs are spatially varying, particularly when the angle of incidence increases. Consequently, these areas experience a noticeable drop in the reconstructed similarity to the original scene, especially in the peripheral field of view.

To achieve both photorealism and consistency, we propose a two-stage lensless reconstruction based on range-null space decomposition~\cite{schwab2019deep}. According to this decomposition, the reconstructed image consists of two components, one from ``range space'', which can be directly calculated from the lensless measurement (\cref{fig:teaser}a), and the other from ``null space'', which are the detailed textures lost during the lensless imaging process. Therefore, the first stage prioritizes data consistency by recovering the ``range space'' component, which is the low-frequency content that matches the ground truth. The second stage focuses on photorealism by adding more high-frequency details from the ``null space'' while maintaining the consistency established in the first stage.

In the first stage, to improve consistency, we propose a novel spatially varying deconvolution to reconstruct the ``range space'' content.
Unlike traditional methods that assume a shift-invariant PSF, our approach automatically adapts to spatial variations in the PSF across the camera's field of view in a data-driven manner. This innovation more accurately models the forward imaging process, leading to better reconstruction of structural integrity and detail in the low-frequency components. \cref{fig:svd_comparson} shows improvement with our method (right) against those using a single kernel for deconvolution(left).


In the second stage, to improve photorealism, we integrate a generative prior using a pre-trained diffusion model, to insert realistic details into the first-stage output. This model specifically targets the recovery of high-frequency details lost in the lensless imaging process. We condition the diffusion model on the low-frequency content reconstructed in the first stage, guiding it to incorporate missing high-frequency elements from the ``null space''. Our supervised approach ensures the final images are consistent with actual measurements and also achieve photorealistic quality.

Combining these two stages, our lensless reconstruction achieves a good balance between data consistency and visual quality in the reconstructed images, surpassing existing methods, as shown in~\cref{fig:teaser}e. 
To validate this, we compare our method with others on two types of lensless cameras: PhlatCam \cite{khan2020flatnet} and DiffuserCam \cite{monakhova2019learned}, using metrics that assess both fidelity and visual quality. The qualitative comparison on the PhlatCam dataset, demonstrating improvements in both aspects, is shown in \cref{fig:radar}, highlighting the superiority of our technique over current methods.

\vspace{-0.5em}
\section{Related Work}
\vspace{-0.3em}

\textbf{Lensless imaging} is traditionally addressed by solving regularized least squares problems for convolutional imaging models, often incorporating sparsity constraints like those in the gradient or frequency domain \cite{antipa2018diffusercam,boominathan2020phlatcam,li2013efficient,reddy2011p2c2}.
Deep learning has revolutionized lensless imaging by enabling learnable deconvolution parameters and perception enhancement through image-to-image networks.
Recent innovations include deep unrolled techniques \cite{kingshott2022unrolled,monakhova2019learned, zeng2021robust}, alongside feed-forward deconvolution methods in image space \cite{khan2020flatnet} or in feature space \cite{li2023mwdns}.
However, these methods generally assume a constant point spread function in the imaging process. Our proposal introduces a spatially varying deconvolution approach to overcome this limitation and achieve more precise image reconstruction.

\textbf{Spatially-varying deconvolution} has been a well-studied area due to the prevalence of imaging systems with PSFs that vary across the FoV. However, these methods \cite{arigovindan2010parallel, kuo2020chip, maalouf2011fluorescence,patwary2015image,yanny2020miniscope3d} are often slow, computationally intensive, and result in poor image quality, especially in complex systems.
Recently, MultiWienerNet \cite{yanny2022deep} introduced a deep learning approach for fast, spatially varying deconvolution, but it requires tedious multi-location PSF calibrations. In contrast, our novel spatially varying deconvolution method, designed specifically for lensless imaging, leverages a single initial PSF and automatically learns variations across the image, eliminating the need for extensive calibration.

 \textbf{Inverse imaging with diffusion models} can be categorized as supervised or zero-shot. Supervised methods train a conditional diffusion model \cite{choi2021ilvr,ho2022classifier,ho2020denoising,rombach2022high,zhang2023adding} with paired images to bridge the gap between input and desired outputs, leveraging generative priors for inverse imaging \cite{saharia2022image,wang2023exploring} and controlled generation \cite{mou2024t2i, saharia2022palette}. On the other hand, zero-shot methods \cite{song2019generative} employ guidance to address a wide range of general inverse problems \cite{chung2022improving,song2021solving}. These techniques typically rely on predefined conditions for guidance. Specific models like DDRM \cite{kawar2022denoising} and DDNM \cite{wang2022zero} focus on mathematical decompositions to improve the diffusion process. Our work integrates the strengths of both approaches, combining supervised fine-tuning with a theoretical framework based on range-null space decomposition, aiming for a robust solution to inverse imaging challenges.

\vspace{-0.5em}
\section{Preliminary}
\vspace{-0.3em}

\subsection{Range-Null Space Decomposition}\label{sec:RNSD}
\vspace{-0.5em}

The lensless imaging process can be formulated as a linear transformation~\cite{goodman2005introduction}. Specifically, given a scene plane $\mathbf{x} \in \mathbb{R}^{M^2}$, the sensor measurements $\mathbf{\hat y} \in \mathbb{R}^{N^2}$ is expressed as $\mathbf{\hat y} = \mathbf{Ax} + \mathbf{n}$. Here, $\mathbf{A} \in \mathbb{R}^{N^2 \times M^2}$ is the transfer matrix of the lensless imaging system, and $\mathbf{n}$ is the sensor noise.
The transfer matrix $\mathbf{A}$ essentially encodes how light from each point on the scene plane contributes to each sensor pixel. Given a calibrated camera system, the transfer matrix $\mathbf{A}$ is known and the objective of lensless reconstruction is to recover the scene plane image $\mathbf{x}$ (like \cref{fig:teaser}f) from the measurement (like \cref{fig:teaser}a). For simplicity, below we consider a noise-free case as $\mathbf{y} = \mathbf{Ax}$.

We then introduce range-null space decomposition. Let $\mathbf{A}^{\dag} \in \mathbb{R}^{M^2 \times N^2}$ be the pseudo-inverse of the linear matrix $\mathbf{A}$, which satisfies $\mathbf{A}\mathbf{A}^{\dag}\mathbf{A} \equiv \mathbf{A}$. The operation $\mathbf{A}^{\dag}\mathbf{A}$ can be interpreted as a projection onto the range space of $\mathbf{A}$ because for any sample $\mathbf{x}$, we have $\mathbf{A}\mathbf{A}^{\dag}\mathbf{A}\mathbf{x} = \mathbf{A}\mathbf{x}$. In contrast, the operation $(\mathbf{I} - \mathbf{A}^{\dag}\mathbf{A})$ acts as a projection onto the null space of $\mathbf{A}$, given that $\mathbf{A}(\mathbf{I} - \mathbf{A}^{\dag}\mathbf{A})\mathbf{x} = \mathbf{0}$. Therefore, any sample $\mathbf{x}$ can be decomposed into two orthogonal components: one that lies in the range space of $\mathbf{A}$, and the other in the null space of $\mathbf{A}$. Mathematically, this decomposition is:
\vspace{-0.4em}
\begin{equation}
    \mathbf{x} \equiv \mathbf{A}^{\dag}\mathbf{A}\mathbf{x} + (\mathbf{I} - \mathbf{A}^{\dag}\mathbf{A})\mathbf{x},
\end{equation}
\vspace{-0.6em}
where $\mathbf{A}^{\dag}\mathbf{A}\mathbf{x}$ is the range space component, and $(\mathbf{I} - \mathbf{A}^{\dag}\mathbf{A})\mathbf{x}$ is the null space component. 

Through this decomposition, we derive two distinct elements indicative of consistency and photorealism.
The ``range space'' term $\mathbf{A}^{\dag}\mathbf{A}\mathbf{x}$ ensures consistency, as its multiplication by $\mathbf{A}$ yields the measurement $\mathbf{y}$, fulfilling the consistency condition $\mathbf{A}\mathbf{x} = \mathbf{y}$. Conversely, the ``null space'' term $(\mathbf{I} - \mathbf{A}^{\dag}\mathbf{A})\mathbf{x}$ ensures that the reconstructed image $\mathbf{\hat{x}}$ aligns with natural image statistics.

\vspace{-0.5em}
\subsection{Mismatch in Convolutional Lensless Imaging Model}\label{sec:mismatch}
\vspace{-0.3em}

In lensless literature, most researchers~\cite{antipa2018diffusercam, boominathan2020phlatcam,khan2020flatnet,monakhova2019learned} simplify the imaging process as a convolution, because the full transfer matrix $\mathbf{A} \in \mathbb{R}^{N^2 \times M^2}$ is too large to compute. Specifically, the measurement $\mathbf{y}$ is obtained as $\mathbf{y} = \mathbf{h * x} $, where $\mathbf{h}$ is the point spread function (PSF) of the lensless system and $*$ denotes the convolution. Most previous reconstruction algorithms are based on this assumption.

\begin{figure}[t]
\centering
\vspace{-10pt}
\includegraphics[width=\linewidth]{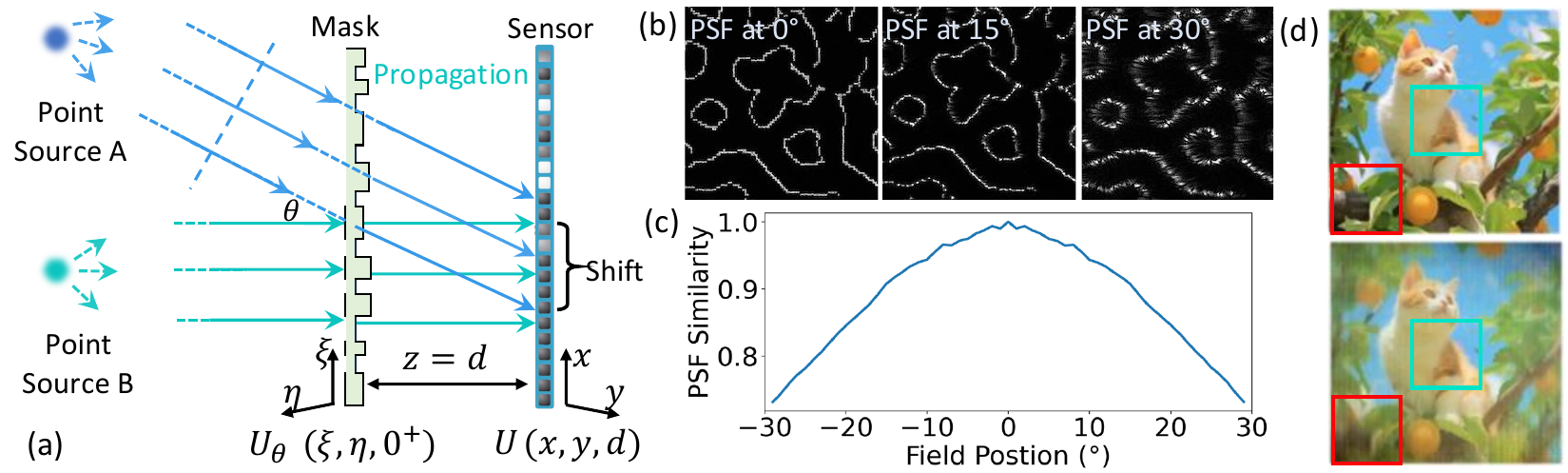}
\caption{Characterization of PSFs in Lensless Camera. (a) Illustration of light propagation in the lensless camera: two point sources A and B at infinity emitting parallel light beams. Source A emits at angle $\theta$ relative to the optical axis, causing a PSF shift on the sensor plane This PSF shift depends on both the incident angle $\theta$ and the distance $d$ between the sensor and the mask. (b) Simulated PSFs for light sources at angles of 0°, 15°, and 30°. (c) Inner product similarity between the on-axis PSF and off-axis PSFs at different field positions. (d) Reconstruction using PSF at 0°, degradation is more significant at the periphery (\textcolor{red}{red} box) than the center (\textcolor{darkgreen}{green} box).}
\label{fig:svpsf}
\vspace{-10pt}
\end{figure}

However, the real lensless imaging model is not simply a spatially invariant convolution as shown in \cref{fig:svpsf}a. Consider a lensless camera consisting of a mask placed at the longitudinal position $z=0$ and a sensor placed at $z=d$. Let $U(x,y,z)$ be a complex scalar wave field,  a complex function of transverse coordinates $x,y$, and longitudinal position $z$. Denote the wave field immediately after propagating through the mask from a point source at infinity with an incident angle $\theta$ as $U_\theta(\xi,\eta,0^+)$. Using the Huygens-Fresnel principle \cite{goodman2005introduction}, the intensity pattern $p_\theta(x,y)$ captured by the sensor is:
\vspace{-.5em}
\begin{equation}
    p_\theta(x,y) = |U_\theta(x,y,d)|^2 =\left|\frac{d}{j\lambda r^2}\iint U_\theta(\xi,\eta,0^+)\exp(jkr)d\xi d\eta\right|^2,
    \label{eq:huygens}
\end{equation}
where the distance $r$ is given by $r = \sqrt{d^2 + (x - \xi)^2 + (y - \eta)^2}$ and $\lambda$ is the wavelength of light.

Therefore, according to Eqn.~\ref{eq:huygens}, the spatially invariant convolutional imaging model only satisfies under the Fresnel approximation~\cite{boominathan2020phlatcam,goodman2005introduction}:
\begin{equation}
    r = \sqrt{d^2 + (x - \xi)^2 + (y - \eta)^2}  \approx d + (x - \xi)^2/(2d) + (y - \eta)^2/(2d).
\end{equation}
This approximation is valid only when the distance $d$ between the lensless mask and the sensor is large enough to satisfy $d \gg \sqrt{(x-\xi)^2 + (y - \eta)^2}$. However, most lensless masks are very close to the sensor ($d = 2\rm{mm}$ in a typical lensless camera), breaking this assumption.

This mismatch can lead to inaccuracies in the convolutional imaging model. To demonstrate that, we simulate the PSF of a typical lensless camera (Phlatcam \cite{boominathan2020phlatcam})
across various incident angles $\theta$ using \cref{eq:huygens}. As shown in \cref{fig:svpsf}b, PSFs at different angles (0°, 15°, and 30°) are visually different. Quantitatively, \cref{fig:svpsf}c shows the similarity between PSFs at the center and -30° drops from 1.0 to 0.7.

To further show the consequences of this mismatch, we simulate lensless capture from a clean scene image (\cref{fig:svpsf}d top), with the accurate spatially-varying PSF based on ~\cref{eq:huygens}, but solve the inverse imaging problem using a simple convolutional model using the PSF at the center. \cref{fig:svpsf}d bottom shows the result, and there are more artifacts at the boundary (red box) than at the center (green box). This is because the mismatch between the actual PSF and the single PSF used for deconvolution is more significant at the periphery field of view.

\vspace{-5pt}
\section{Method}
\vspace{-5pt}

\begin{figure}[t]
\vspace{-5pt}
\centering
\includegraphics[width=\linewidth]{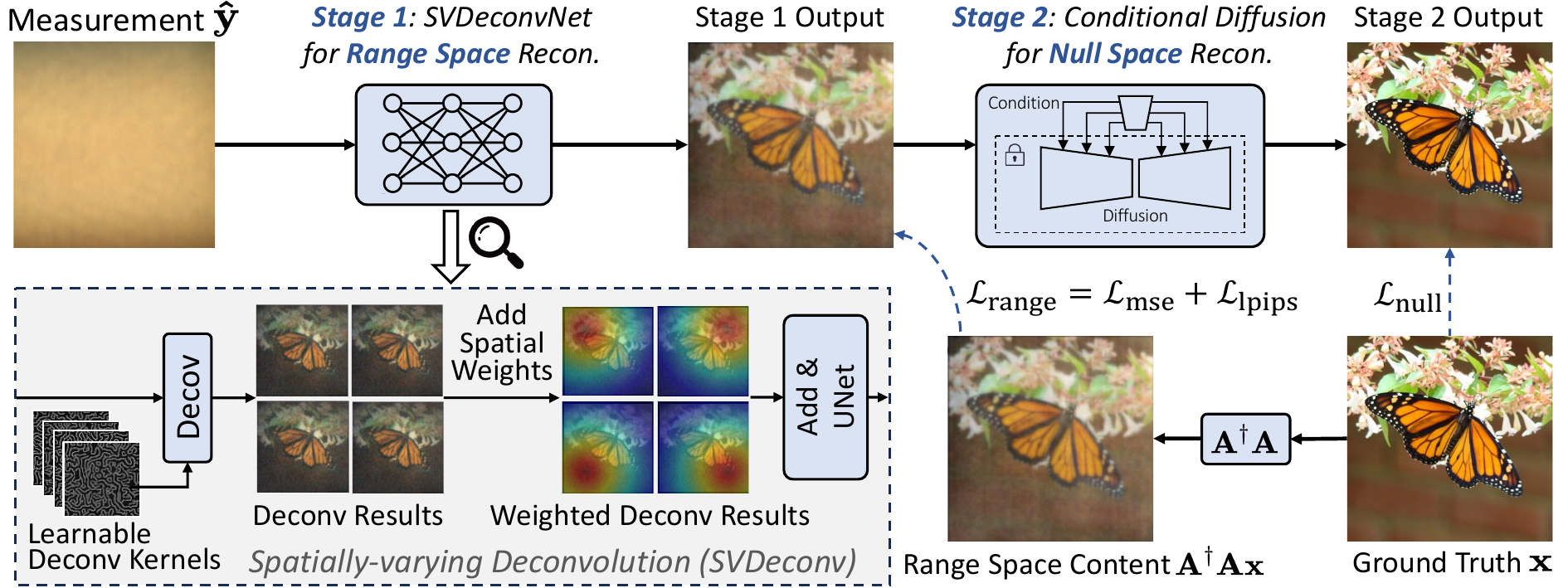}
\vspace{-15pt}
\caption{
    System Overview. The two-stage pipeline begins with a spatially varying deconvolution network mapping lensless measurements to range space. Then a conditional diffusion model for null space recovery refines details using the first stage output, achieving the final reconstruction. 
}\label{fig:method}
\vspace{-15pt}
\end{figure}

In this section, we introduce \textbf{PhoCoLens}, a method for achieving both \textbf{Pho}torealistic and \textbf{Co}nsistent reconstruction in \textbf{Lens}less imaging. As illustrated in ~\cref{fig:method}, PhoCoLens consists of two main stages: range space reconstruction and null space recovery. We will first introduce the whole framework and then provide a detailed explanation of each stage.

Given an input lensless measurement $\mathbf{\hat y} =  \mathbf{A}\mathbf{x} + \mathbf{n}$, where $\mathbf{y} = \mathbf{A}\mathbf{x}$ is the noise-free part of the lensless measurement, $\mathbf{A}$ is the transfer matrix of the lensless camera, $\mathbf{x}$ is the original scene we aim to reconstruct, and $\mathbf{n}$ is the sensor noise.
The objective in lensless image reconstruction is to recover a photo-realistic image $\mathbf{\hat{x}}$ which satisfies $\mathbf{A}\mathbf{\hat x} = \mathbf{y}$.

Inspired by the range null space decomposition in \cref{sec:RNSD}, we can decompose any potential solution $\mathbf{\hat x}$ into two orthogonal components: the range space component that maintains consistency and null space component that maximizes photorealism.
Formally, this decomposition is given by $\mathbf{\hat x} = \mathbf{A}^{\dag}\mathbf{A}\mathbf{x} + (\mathbf{I} - \mathbf{A}^{\dag}\mathbf{A})\mathbf{\bar x}$, where $\mathbf{A}^{\dag}\mathbf{A}\mathbf{x}$ is the range space component and $(\mathbf{I} - \mathbf{A}^{\dag}\mathbf{A})\mathbf{\bar x}$ is the null space component. Note that any choice of $\mathbf{\bar x}$ satisfies the equation $\mathbf{A}\mathbf{\hat{x}} = \mathbf{y}$, as $\mathbf{A}(\mathbf{I} - \mathbf{A}^{\dag}\mathbf{A})\mathbf{\bar{x}} = \mathbf{0}$.

With this decomposition, we reconstruct the range space and null space components in two stages, as illustrated in ~\cref{fig:method}.
In the first stage, we recover the range space content represented by $\mathbf{A}^{\dagger} \mathbf{A} \mathbf{x}$ from the noisy lensless measurement $\mathbf{\hat y}$. Specifically, we propose \textbf{SVDeconv}, a physics-inspired \textbf{S}patially-\textbf{V}arying \textbf{Deconv}olution Network that learns to reconstruct the range space content effectively.
The second stage focuses on adding the null space content represented by $(\mathbf{I} - \mathbf{A}^{\dagger} \mathbf{A}) \bar{\mathbf{x}}$. For this purpose, we introduce \emph{null space diffusion}, a conditional diffusion model designed for null space recovery. By conditioning on the range space reconstruction obtained from the first stage, the null space diffusion enhances these images by incorporating the null space content, thereby rendering them with a more realistic appearance. Below we introduce each stage.

\vspace{-6pt}
\subsection{Spatially-varying Deconvolution Network} \label{sec:svdeconv}
\vspace{-4pt}

In the first stage, we use SVDeconv, a novel network designed to invert the forward lensless imaging model and reconstruct the range space content $\mathbf{A}^{\dagger} \mathbf{A} \mathbf{x}$ from a lensless measurement $\mathbf{\hat y} = \mathbf{A}\mathbf{x} + \mathbf{n}$ with noise $\mathbf{n}$. SVDeconv is composed of two main components: a differentiable multi-kernel deconvolution layer and a refinement U-Net, as depicted in the gray box at the bottom left corner of \cref{fig:method}.

\vspace{-1pt}
Traditional lensless imaging methods often employ a single PSF to model the imaging system, which is inaccurate for large field of view (FoV) scenarios as analyzed in ~\cref{sec:mismatch}. To address this, SVDeconv utilizes a set of learnable $K \times K$ PSF kernels, accounting for spatial variations across the field of view. Specifically, we partition the target image region into a grid of $K \times K$ segments, and for each, we apply an individual PSF kernel. This process results in $K \times K$ deconvolution operations, producing $K \times K$ intermediate deconvolved images, as shown in \cref{fig:method}.
The multi-kernel deconvolution is mathematically described as a Hadamard product in the Fourier domain:
 \vspace*{-4pt}
\begin{equation}
    \mathbf{x}^{(i)}_{\mathrm {de}} = \mathcal{F}^{-1}(\mathcal{F}(\mathbf{p}^{(i)}) \odot \mathcal{F}(\mathbf{\mathbf{\hat y}}) ), i = 1,2,..., K \times K,
\end{equation}
where $\mathbf{x}^{(i)}_{\mathrm {de}}$ is the $i$-th deconvolution results of the $i$-th learnable PSF kernel $\mathbf{p}^{(i)}$, $\mathcal{F}(.)$ and $\mathcal{F}^{-1}(.)$ are the DFT and the Inverse DFT operations, and $\odot$ denotes the Hadamard product.

From this deconvolution operation, we obtain $K \times K$ intermediate deconvolved images. Assuming that each image accurately reconstructs a specific region of the target image corresponding to its PSF's field point, we propose to integrate these images into a single, unified intermediate image using an innovative interpolation method.  The interpolation is expressed as:
\vspace{-5pt}
\begin{equation}
       \mathbf{x}_{\text{int}}(u,v) = \sum_{i=1}^{K^2}w_i(u,v) \mathbf{x}^{(i)}_{\text{de}}(u,v), 
\end{equation}
where $u,v$ are coordinates. The weight $w_i(u,v)$ for each deconvolved image $\mathbf{x}^{(i)}_{\text{de}}$ is inversely proportional to the distance between the point $(u,v)$ and the focal center of each corresponding PSF, formally defined as:
\vspace*{-4pt}
\begin{equation}
    w_i(u,v) = \frac{d_i^{-\frac{1}{2}}(u,v)}{\sum_{j=1}^{K^2}d_j^{-\frac{1}{2}}(u,v)}, \mathrm{with} \enspace  d_i(u,v) = (u-u_i)^2 + (v-v_i)^2.
\end{equation}
In this way, it ensures that $w_i(u,v)$ normalizes the contribution of each deconvolved image based on inverse Euclidean distance to each point $(u_i, v_i)$, which represents the center of the FoV corresponding to the $i$-th learnable PSF kernel $\mathbf{p}^{(i)}$. This weighted sum effectively interpolates the intermediate images to form a unified representation with enhanced clarity and detail across the entire FoV.

One advantage of this weight design is that it significantly simplifies the calibration process. Previous multi-kernel methods, such as \cite{yanny2022deep}, often require multiple calibrated PSFs at different focal centers, leading to a time-consuming calibration process. On the contrary, our approach only requires one calibrated PSF to initialize all kernels. This is because we pre-define the center of the PSF and allow the model to learn its variations automatically.

After the interpolation, the intermediate image $\mathbf{x}_{\text{int}}$ is fed into the refinement U-Net~\cite{ronneberger2015u}, removing noise and artifacts in the reconstruction to approximate the range space content $\mathbf{A}^{\dagger} \mathbf{A} \mathbf{x}$. We employ a combination of MSE loss and LPIPS loss ~\cite{zhang2018unreasonable} to train both the learnable PSFs and U-Net.

\vspace{-5pt}
\subsection{Null Space Content Recovery} \label{sec:NSCR}
\vspace{-5pt}

With the reconstructed range space content, we aim to recover a final image that maintains consistency and improves photorealism.
To ensure consistency, the difference between the final image and the range space content (residual content) should reside in the null space, which ensures the final image aligns with the original lensless measurement.
To improve photorealism, the combination of the null space and range space content should appear as a realistic real-world image.

To achieve both objectives, we propose null-space diffusion, which takes the reconstructed range space content as the condition and generates an image that adheres to both constraints. It ensures the residual content lies in the null space while maintaining its ability to produce realistic images.

To ensure the outputs from null-space diffusion meet the consistency requirement, we train the model such that the residual content of these samples falls within the null space. Mathematically, if we denote the output as $\mathbf{\widetilde{x}}$, it should satisfy the following equation when conditioned by $\mathbf{A}^{\dagger}\mathbf{A}\mathbf{x}$:
\vspace{-5pt}
\begin{equation} \label{eq:consistency}
    \mathbf{A}(\mathbf{\widetilde{x}} - \mathbf{A}^{\dagger}\mathbf{A}\mathbf{x}) = \mathbf{0} \quad \Longrightarrow \quad \mathbf{A}\mathbf{\widetilde{x}} - \mathbf{A}\mathbf{x} = \mathbf{0} \quad \Longrightarrow \quad \mathbf{A}(\mathbf{\widetilde{x}} - \mathbf{x}) = \mathbf{0}.
\end{equation}
Recall that $\mathbf{A}$ is the transfer matrix of the lensless camera and $\mathbf{A}^{\dagger}$ is its pseudo-inverse. This equation indicates that applying the operator $\mathbf{A}$ to the difference between the generated sample $\mathbf{\widetilde{x}}$ and the original sample $\mathbf{x}$ results in zero. This ensures that the residual content resides in the null space of $\mathbf{A}$.

Based on this, we design the null space diffusion as follows. Given an image $\mathbf{x}$ and its corresponding range space content $\mathbf{c} =  \mathbf{A}^{\dagger}\mathbf{A}\mathbf{x}$, we add noise progressively to the image resulting in a noisy image $\mathbf{x}_t$, where $t$ is the noise addition iteration. Similar to other conditional diffusion models \cite{zhang2023adding}, we train a network $\epsilon_\theta$ to predict the noise added on the noisy image $\mathbf{x}_t$ with the optimization objective:
\vspace{-3pt}
\begin{equation}\label{eq:diffusion}
  \min_\theta \mathcal{L}_{\rm{null}} = \min_\theta 
 \mathbb E_{\mathbf{x},t,\mathbf{c},\epsilon \sim \mathcal{N}(0,1)}[||\mathbf{A}(\epsilon_\theta(\mathbf{x}_t,t,\mathbf{c}) - \epsilon)||^2_2],
\end{equation}
\vspace{-3pt}
which is derived to align with the goal in \cref{eq:consistency}. More details are in the Appendix. 

Additionally, to ensure the null-space diffusion produces realistic images according to given conditions, we utilize a pre-trained diffusion model such as Stable Diffusion \cite{rombach2022high} with its weights frozen. 
We focus on training supplementary conditioning modules to guide the generative process using range space conditions while preserving its powerful ability to generate realistic images. Specifically, we follow the StableSR \cite{wang2023exploiting} structure, which involves using a conditional encoder to extract multi-scale features from the range space condition and then using them to modulate the intermediate feature maps of the residual blocks in the diffusion model. This modulation can align the generated images with the range space conditions, enhancing the fidelity and realism of output images.

\vspace{-1em}
\section{Experiments} \label{sec:exp}
\vspace{-0.7em}
In this section, we assess the performance of our proposed method using two lensless imaging datasets, collected in real-world environments by two different kinds of lensless cameras: PhlatCam ~\cite{boominathan2020phlatcam} and DiffuserCam ~\cite{antipa2018diffusercam}. PhlatCam employs a phase mask, while DiffuserCam utilizes a diffuser for its lensless mask. We compare our approach with other methods and conduct a comprehensive ablation study to evaluate the effectiveness of our design.

\vspace{-0.5em}
\subsection{Dataset and Metrics}
\vspace{-0.5em}

The \textit{PhlatCam} dataset~\cite{khan2020flatnet} contains 10,000 images across 1,000 classes resized to 384×384 pixels. Images are displayed and captured using a lensless PhlatCam \cite{boominathan2020phlatcam}, generating images at 1280 × 1480 pixels. We use 990 classes for training and 10 classes for testing, following the original protocol.

The \textit{DiffuserCam} dataset \cite{monakhova2019learned} contains 25,000 paired images captured simultaneously using a standard lensed camera (ground truth) and a mask-based lensless camera DiffuserCam \cite{antipa2018diffusercam}. These pairs are split into 24,000 images for training and 1,000 for testing. Both cameras utilize sensors with a native resolution of 1080×1920 pixels, which are first downsampled to 270×480 pixels and then further cropped to a final resolution of 210×380 pixels for proper display.

To evaluate both consistency (fidelity) and photorealism (visual quality), we employ two metric sets:
\vspace{-0.7em}
\begin{itemize}
\setlength{\itemsep}{0.0em}
\setlength{\parsep}{0.0em} 
\item \noindent \textbf{Full-reference metrics} to evaluate the consistency. Three full-reference metrics, PSNR, SSIM \cite{wang2004image}, and LPIPS \cite{zhang2018unreasonable}, are used to evaluate the distance between the network output and ground truth, which indicates the fidelity of the reconstruction.
\item \textbf{Non-reference metrics} to evaluate the photorealism. Three non-reference metrics, ManIQA \cite{yang2022maniqa}, ClipIQA \cite{wang2023exploring}, and MUSIQ \cite{ke2021musiq}, are used to evaluate the visual quality of reconstructed images.
\end{itemize}

\vspace{-1em}
\subsection{Implementation Details} \label{sec:implement}
\vspace{-0.3em}

For SVDeconv, we utilize $3\times3$ PSF kernels for deconvolution. We initialize the 9 kernels using a single calibrated PSF from both the DiffuserCam and PhlatCam datasets. The U-Net component in SVDeconv is adapted from the U-Net used in Le-ADMM-U for training on the DiffuserCam, and from the U-Net in FlatNet-gen for the PhlatCam. We train the network for 100 epochs with a batch size of 5, using the Adam optimizer ~\cite{kingma2014adam}. The learning rate is set to 3e-5 for U-Net training in both datasets. For deconvolution kernel training, the learning rate is set at 4e-9 for PhlatCam and 3e-5 for DiffuserCam. We set the MSE and LPIPS loss weights to be 1 and 0.05, respectively. In the DiffuserCam dataset, where the measurements are cropped by the sensor's limited size, we employ replicate padding \cite{khan2020flatnet} which expands the width and height of the measurements by a factor of two.

For null-space diffusion, we use the range space content reconstructed by SVDeconv as input conditions. We train it for 200 epochs, following the training and inference settings used in StableSR\cite{wang2023exploring}.

In the first stage, SVDeconv is trained using range-space content derived from ground truth images. SVDeconv consists of two main parameter components: a learnable deconvolution kernel initialized with known PSFs, and a U-Net initialized with standard weights without pretraining. Once trained, SVDeconv processes input lensless measurements to estimate the range-space content of training samples. Subsequently, we use this estimated range-space content as input conditions for fine-tuning via null-space diffusion. During diffusion fine-tuning, we utilize a pre-trained diffusion model with frozen weights. We only train the supplementary conditioning modules like StableSR [35], to guide the reconstruction process effectively.

\begin{table}[tbp]
  \centering
  \caption{Performance comparison of methods on DiffuserCam and PhlatCam}\label{tab:comparison}
{\setlength\tabcolsep{2pt}
   \fontsize{7.8}{10}\selectfont
  \begin{tabular}{@{}lcccccc|cccccc@{}}
    \toprule
    Dataset & \multicolumn{6}{c|}{PhlatCam} & \multicolumn{6}{c}{DiffuserCam} \\
    \cmidrule(r){1-7} \cmidrule(l){8-13}
    Metrics & PSNR & SSIM & LPIPS$\downarrow$ & ManIQA & ClipIQA & MUSIQ & PSNR & SSIM & LPIPS$\downarrow$ & ManIQA & ClipIQA & MUSIQ \\
    \midrule
    \textit{Ground Truth} & --- & --- & --- &  \textit{0.431} &  \textit{0.583} &  \textit{63.40} & --- & --- & --- & \textit{0.160}  & \textit{0.333}   & \textit{31.21} \\
    WienerDeconv \cite{wiener1949extrapolation} & 12.19 & 0.270 & 0.922 & 0.111 & 0.149 & 15.47 & 10.59 & 0.275 & 0.843 & 0.184 & 0.180 & 19.43 \\
    ADMM \cite{boyd2011distributed} & 13.45 & 0.301 & 0.877 & 0.132 & 0.159 & 16.45 & 12.87 & 0.305 & 0.705 & 0.141 & 0.136 & 16.51 \\
    Le-ADMM-U \cite{monakhova2019learned} & 20.12 & 0.515 & 0.405 & 0.138 & 0.180 & 24.64 & 22.35 & 0.668 & 0.253 & 0.110 & 0.185 & 20.16 \\
    MMCN \cite{zeng2021robust} & 20.39 & 0.524 & \underline{0.346} & 0.145 & 0.206 & 27.65 & 24.09  & 0.744 & \underline{0.238} & 0.121 & 0.183 & 21.47 \\
    UPDN \cite{kingshott2022unrolled} & 20.48 & 0.533 & 0.352 & 0.158 & 0.215 & 29.32 & \textbf{24.67} & \underline{0.747} & 0.256 & 0.139 & 0.178 & 22.56 \\
    FlatNet-gen \cite{khan2020flatnet} & \underline{20.53} & \underline{0.549} & 0.375 & 0.203 & 0.287 & 44.94 & 21.43 & 0.696 & 0.254 & 0.134 & 0.244 & 23.15 \\
    DDNM+\cite{wang2022zero} & 15.22 & 0.485 &  0.623 &  0.297 & 0.412 & 45.32 & 18.36 & 0.539 & 0.516 & 0.162 & 0.201 & 27.83 \\
    FlatNet+DiffBIR \cite{lin2023diffbir} & 19.96 & 0.544 & 0.391 & \underline{0.335} & \underline{0.523} & \underline{57.13} & 18.97 & 0.517 & 0.505 & \textbf{0.312 }& \textbf{0.548} & \textbf{51.20} \\
    \textbf{PhoCoLens(Ours)} & \textbf{22.07} & \textbf{0.601} & \textbf{0.215} & \textbf{0.357} & \textbf{0.565} & \textbf{62.20} & \underline{24.12} & \textbf{0.748} & \textbf{0.161} & \underline{0.172} & \underline{0.339} & \underline{32.84} \\

    \bottomrule
  \end{tabular}
  }
\end{table}

\begin{figure}[thbp]
\centering
\includegraphics[width=\linewidth]{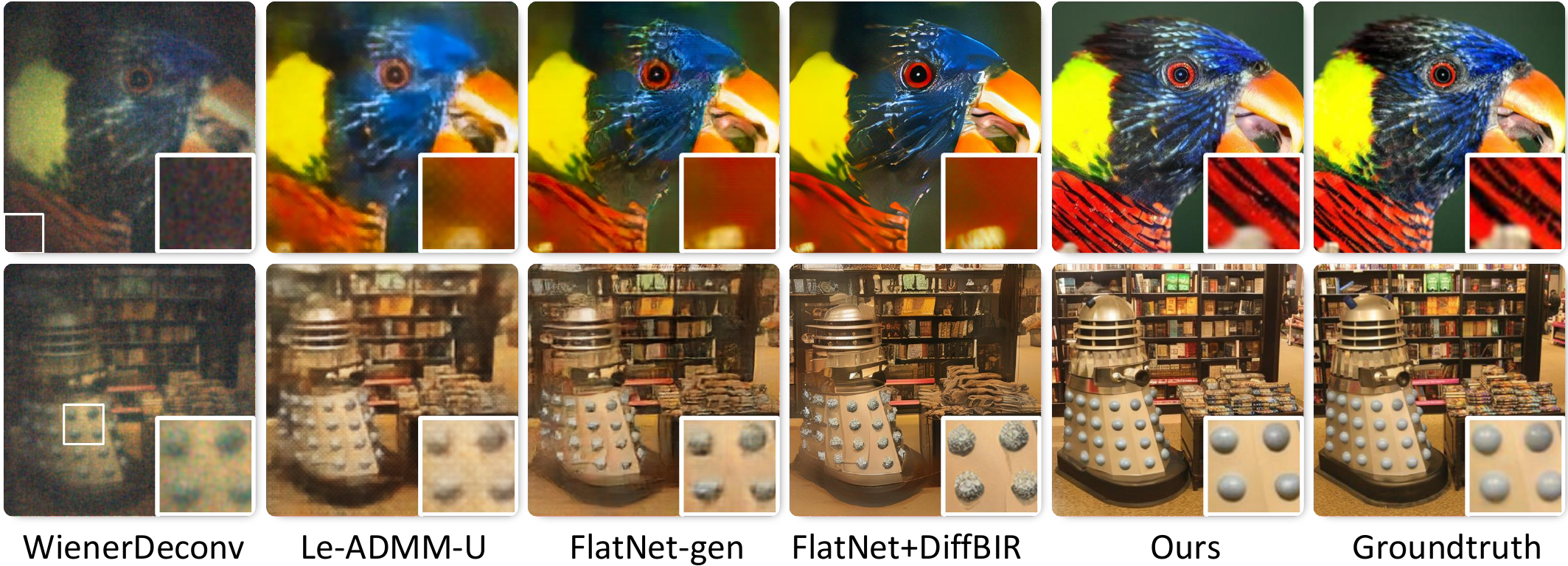}
\caption{
    Qualitative comparison between our method and others on the PhlatCam dataset. 
}
\label{fig:main_result}
\vspace{5pt}
\includegraphics[width=\linewidth]{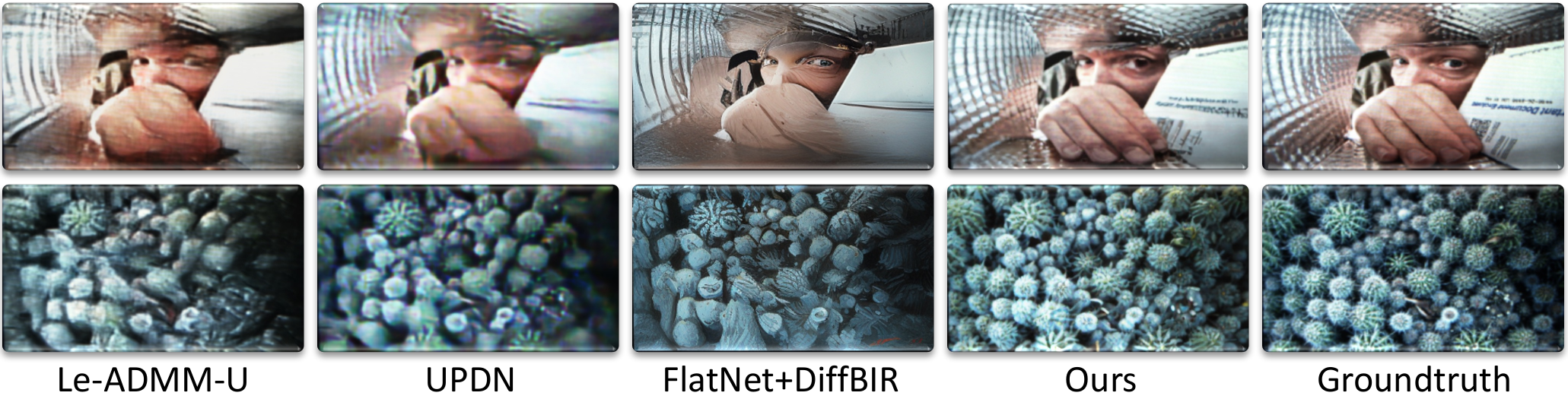}
\caption{
    Qualitative comparison between our method and others on the DiffuserCam dataset. 
\label{fig:main_result2}
}
\vspace{-20pt}
\end{figure}

\vspace{-1.2em}
\subsection{Comparison with Other Approaches}
\vspace{-0.3em}

We evaluate the performance of our proposed method by comparing it against both traditional and learning-based approaches, using qualitative and quantitative measures on the two datasets.

We compare traditional methods like Tikhonov regularized reconstruction in the Fourier domain (WienerDeconv \cite{wiener1949extrapolation}) and total variation regularization via ADMM \cite{boyd2011distributed}. Additionally, we evaluate learning-based methods like the unrolled network Le-ADMM-U \cite{monakhova2019learned},  MMCN \cite{zeng2021robust}, and UPDN \cite{kingshott2022unrolled} and the feedforward deconvolution method FlatNet-gen. Additionally, we assess two diffusion-based methods: one uses a pre-trained Stable Diffusion \cite{rombach2022high} with a zero-shot inverse imaging sampling method DDNM+ \cite{wang2022zero}, and another enhances outputs from FlatNet-gen using a pre-trained blind image restoration model DiffBIR \cite{lin2023diffbir}. Quantitative results are summarized in ~\cref{tab:comparison}. Qualitative comparisons for PhlatCam and DiffuserCam are shown in Figures~\ref{fig:main_result} and~\ref{fig:main_result2}, respectively.

On the PhlatCam dataset, our method achieves the best performance on both fidelity (full-reference metrics) and visual quality (non-reference metrics) according to \cref{tab:comparison}, This is further supported by the qualitative results in \cref{fig:main_result}, where our reconstruction are the closest to the ground truth.

For the DiffuserCam dataset, our method achieves superior performance in consistency metrics (full-reference metrics, SSIM and LPIPS). While UPDN achieves a slightly higher PSNR, the visual quality of its outputs suffers from artifacts, as shown in~\cref{fig:main_result2}.
This slight PSNR drop in our method is likely due to artifacts within the DiffuserCam ground truth itself, such as moiré patterns. These artifacts can penalize high-quality images with a lower PSNR, despite their good visual consistency. This is further evidenced by the visual comparison in~\cref{fig:main_result2}.
Conversely, FlatNet+DiffBIR, though achieving higher quality metrics (non-reference metrics, ManIQA, ClipIQA, and MUSIQ), fails to maintain data consistency (low full-reference metrics, PSNR, SSIM, and LPIPS). Also, as shown \cref{fig:main_result2}, FlatNet+DiffBIR sometimes badly distorts the original image content, like transferring a human hand into clothes in the top row.
Overall, our method achieves the best trade-off between the data consistency and the visual quality. And thanks to the introduction of SVDencov, our method can reconstruct the field-of-view periphery better than other methods.

\vspace{-1em}
\subsection{Analysis and Discussion} \label{sec:analysis}
\vspace{-0.3em}

\begin{figure}[t]
\begin{minipage}[tbp]{0.59\textwidth}
\centering
\captionof{table}{Comparison of deconvolution methods for range space reconstruction and original content reconstruction.}
\vspace{-5pt}
\label{tab:comparison_deconv}
\resizebox{\textwidth}{!}{
{\setlength\tabcolsep{1pt}
   \fontsize{8.5}{9}\selectfont
\begin{tabular}{@{}clcccccc@{}}
\toprule
Reconstruction & Method          & \multicolumn{3}{c}{PhlatCam} & \multicolumn{3}{c}{DiffuserCam} \\ 
\cmidrule(lr){3-5} \cmidrule(lr){6-8}
\multirow{2}{*}[1em]{Target}         &         & PSNR & SSIM & LPIPS$\downarrow$ & PSNR & SSIM & LPIPS$\downarrow$ \\
\midrule
\multirow{4}{*}{Range Space} 
                      
                      & Le-ADMM-U       & 23.87 & 0.835 & 0.307 & 24.24 & 0.748 & 0.282 \\
                      & SingleDeconv     & 25.61 & 0.874 & 0.263 & 24.94 & 0.799 & 0.230 \\
                      & MultiWienerNet  & 26.43 & 0.880 & 0.256 & 26.32 & 0.823 & 0.189 \\
                      & SVDeconv (Ours)     & \textbf{26.97} & \textbf{0.894} & \textbf{0.253} & \textbf{28.84} & \textbf{0.856} & \textbf{0.144} \\
\midrule
\multirow{4}{*}{Original Content} 
                      & Le-ADMM-U       & 20.12 & 0.515 & 0.405 & 22.35 & 0.668 & 0.253 \\
                      & SingleDeconv     & 20.43 & 0.545 & 0.381 & 21.97 & 0.689 & 0.262 \\
                      & MultiWienerNet  & 20.82 & 0.562 & 0.374 & 23.02 & 0.725 & 0.214 \\
                      & SVDeconv (Ours)     & \textbf{21.01} & \textbf{0.571} & \textbf{0.363} & \textbf{25.55} & \textbf{0.781} & \textbf{0.179} \\
\bottomrule
\end{tabular}
}
}
\end{minipage}
\hfill
\begin{minipage}[tbp]{0.39\textwidth}
    \includegraphics[width=0.96\linewidth]{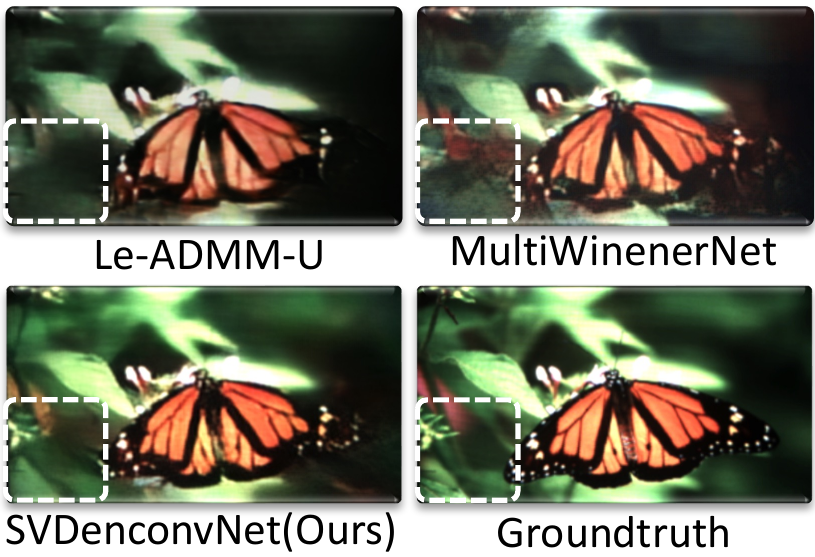}
    \vspace{-5pt}
    \captionof{figure}{Comparison of deconvolution methods on a DiffuserCam example.}\label{fig:svd_result}
\end{minipage}

\vspace{1em}

\begin{minipage}[t]{0.48\textwidth}
\centering
\captionof{table}{ \small Comparison of null space recovery methods}
\label{tab:null_comparison}
\resizebox{\textwidth}{!}{
{
\setlength\tabcolsep{0.9pt}
   \fontsize{10.3}{14}\selectfont
\begin{tabular}{lcccccc}
\toprule
Method       & PSNR & SSIM & LPIPS$\downarrow$ & ManIQA & ClipIQA & MUSIQ \\
\midrule
StableSR \cite{wang2023exploring}    & 14.93 & 0.446 & 0.624       & 0.64  & 0.235   & 33.84  \\
DiffBIR \cite{lin2023diffbir}    & 16.21 & 0.432 & 0.502       & 0.353   &  0.512   & 56.65  \\
DDNM  \cite{wang2022zero}   & 17.62 & 0.539 & 0.517       & 0.148   & 0.201    & 27.82  \\
Ours   &  \textbf{22.07} & \textbf{0.601} & \textbf{0.215} & \textbf{0.357} & \textbf{0.565} & \textbf{62.20}\\
\bottomrule
\end{tabular}
}
}
\end{minipage}
\hfill
\begin{minipage}[t]{0.50\textwidth}
\centering
\captionof{table}{ \small Comparison of different diffusion conditions}
\label{tab:cond_comparison}
\resizebox{\textwidth}{!}{
\setlength\tabcolsep{3pt}
   \fontsize{8.5}{9}\selectfont{
\begin{tabular}{lccc|ccc}
\toprule
\multirow{2}{*}[-0.5em]{Condition} & \multicolumn{3}{c}{PhlatCam} & \multicolumn{3}{c}{DiffuserCam} \\
\cmidrule(lr){2-4} \cmidrule(lr){5-7}
       & PSNR & SSIM & LPIPS$\downarrow$ & PSNR & SSIM & LPIPS$\downarrow$ \\
\midrule
WienerDeconv       & 20.38 & 0.551 & 0.301 & 20.42 & 0.632 & 0.282 \\
FlatNet-gen    & 20.89 & 0.556 & 0.286 & 21.67 & 0.647 & 0.230 \\
SVD-OC    & 21.16 & 0.572 & 0.276 & 22.43 & 0.679 & 0.185 \\
Ours & \textbf{22.07} & \textbf{0.601} & \textbf{0.215} & \textbf{24.12} & \textbf{0.748} & \textbf{0.161} \\
\bottomrule
\end{tabular}
}
}
\end{minipage}

\vspace{1em}

\begin{minipage}[t]{0.47\textwidth}
\centering
\includegraphics[width=\linewidth]{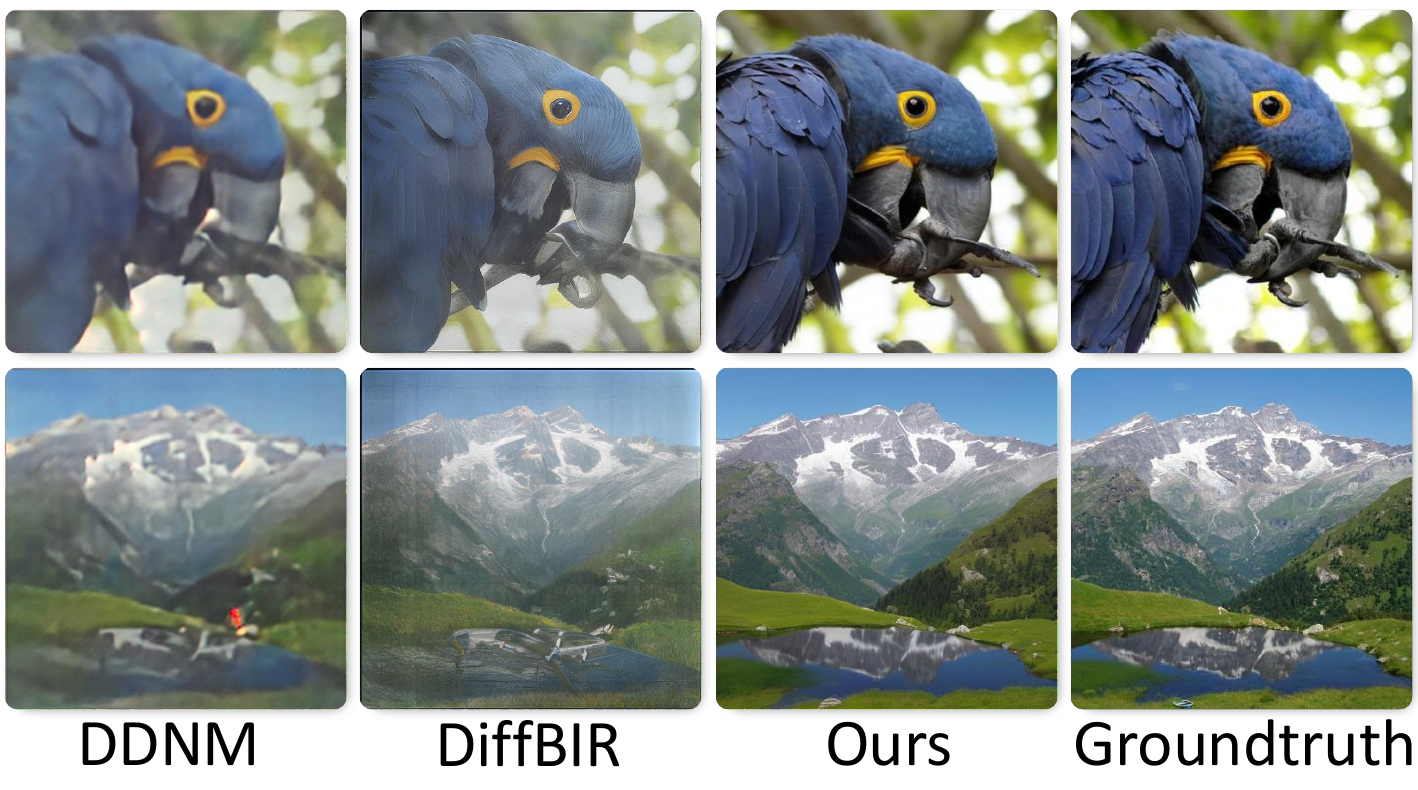}
\vspace{-15pt}
\captionof{figure}{ \small Comparison between our null space diffusion and other methods.}
\label{fig:null_comparison}
\end{minipage}
\hfill
\begin{minipage}[t]{0.47\textwidth}
\centering
\includegraphics[width=\linewidth]{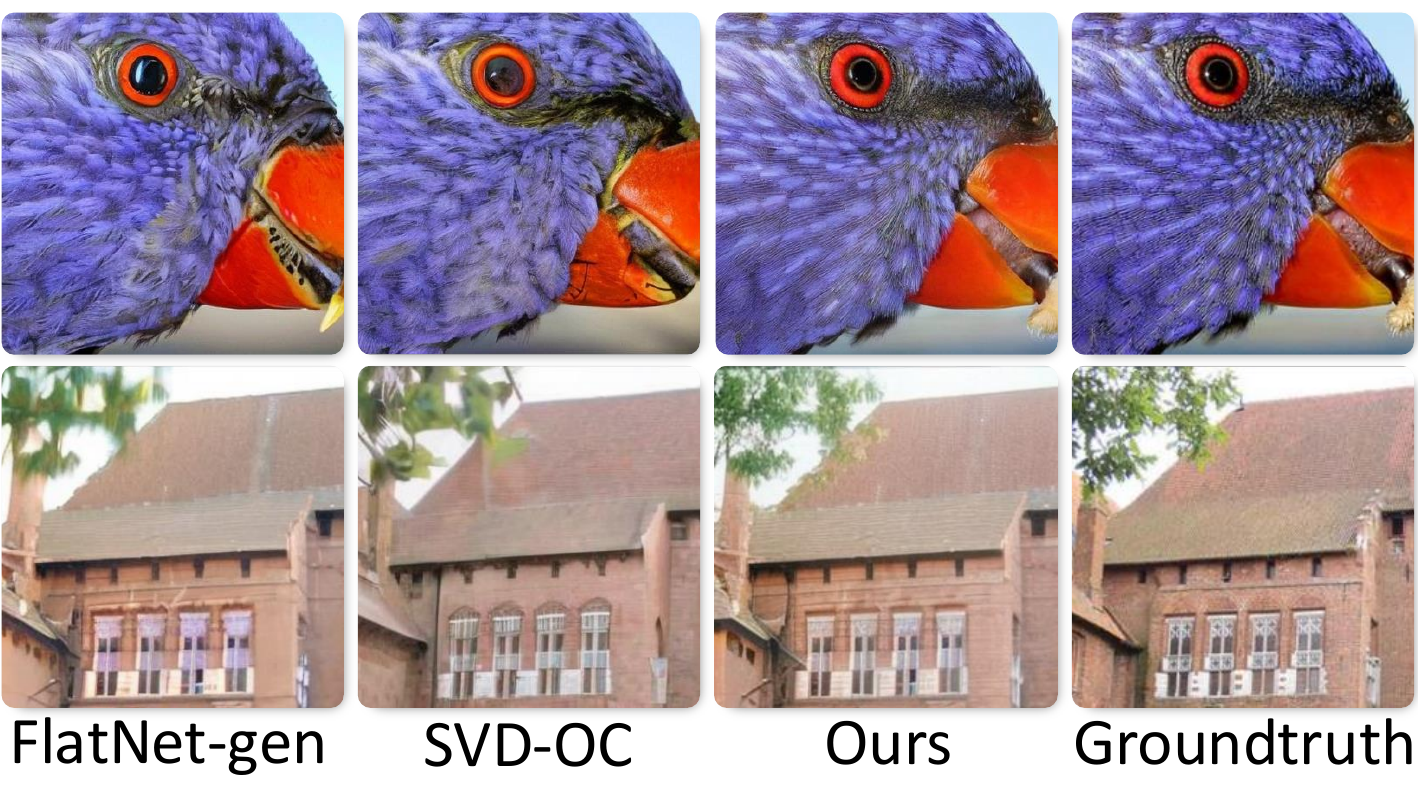}
\vspace{-15pt}
\captionof{figure}{ \small Comparison between different diffusion conditions.}
\label{fig:cond_comparison}
\end{minipage}
\vspace{-2em}
\end{figure}

\textbf{Effectiveness of Spatially-varying Deconvolution.}
To verify the effectiveness of the proposed SVDeconv, we compare it with other learnable deconvolution approaches. Baselines include Le-ADMM-U (a deep unrolled network implementing ADMM), SingleDeconv (using a single deconvolution kernel), and MultiWienerNet \cite{yanny2022deep}(employing multiple deconvolution kernels without weighted interpolation, initialized similarly to our approach with a single PSF). We focus on two types of content reconstruction: range space content reconstruction, where range space content is the target, and original content reconstruction, using the original ground truth as the target.
For a fair comparison, all methods utilize the same U-Net for denoising and perceptual enhancement and the same loss for training. Results in \cref{tab:comparison_deconv} show the efficacy of the proposed spatially-varying deconvolution in the faithful reconstruction of both range space content and original content.
Further, \cref{fig:svd_result} shows our method outperforms others, particularly in reconstructing fine details at the periphery (circled by white dashed boxes).

\textbf{Effectiveness of Null-space Diffusion}.
Following the reconstruction of range space content, we perform null space content recovery conditioned on the previously reconstructed range space content. We compare our null-space diffusion model with other approaches, including conditional diffusion models like DiffBIR \cite{lin2023diffbir} and StableSR \cite{wang2023exploring}, as well as the zero-shot inverse imaging method DDNM \cite{wang2022zero}.
Quantitative results in \cref{tab:null_comparison} and qualitative comparison in \cref{fig:null_comparison} demonstrate that our null-space diffusion performs the best in both consistency with the ground truth and visual quality.

\textbf{Effectiveness of Range Content Conditions}.
One important design of our framework is to use the first stage reconstruction as the condition for the second stage diffusion.
To evaluate the effectiveness of these conditions, we perform an ablation study where we replace our reconstructed range content with the outputs from WienerDeconv, FlatNet-gen, and the SVD-OC (SVDencov trained with original content), and re-train the conditional diffusion models using these conditions.
As evident from \cref{fig:cond_comparison}, our range content condition significantly improves reconstruction consistency compared to others. This is because using output from models like FlatNet-gen introduces additional artifacts that hinder the second stage from recovering the original image.
Furthermore, quantitative analysis in \cref{tab:cond_comparison} confirms this observation. Our method outperforms these conditions in all fidelity metrics. 
\vspace{-1em}
\section{Conclusion and Limitation}\label{sec:conclusion}
\vspace{-0.7em}
We present PhoCoLens, an innovative two-stage approach for achieving photorealistic and consistent reconstructions in lensless imaging. The method leverages the strengths of two complementary approaches: spatially varying deconvolution for the range space ensures consistency, while null-space diffusion for the null space guarantees photorealism. Our experiments on two lensless cameras demonstrate PhoCoLens' effectiveness in reconstructing realistic scenes while preserving fidelity. One limitation of our approach is that the two-stage nature of PhoCoLens and the diffusion model's sampling time hinder its real-time applicability. Additionally, although PhoCoLens achieves the best fidelity, the diffusion model we used may still introduce high-frequency details that deviate from the original scene, particularly for smooth scenes lacking in detail.

\bibliographystyle{plainnat}

\renewcommand\thefigure{A\arabic{figure}}
\renewcommand\thetable{A\arabic{table}}  
\renewcommand\theequation{A\arabic{equation}}
\setcounter{equation}{0}
\setcounter{table}{0}
\setcounter{figure}{0}
\appendix

\newpage
\section*{Appendix}
\section{PhlatCam Simluation}
In \cref{sec:mismatch}, we simulate a PhlatCam to interpret the detail of mismatch in a convolutional lensless imaging model.  Here we explain how to simulate a lensless camera like PhlatCam.

Following previous work~\cite{boominathan2020phlatcam}, our simulation begins by constructing a desired point spread function (PSF) based on the contours generated from random Perlin noise \cite{perlin2002improving}. Given a specific distance between the mask and sensor, we optimize the phase mask to produce a PSF that matches the target. For this optimization, we employ the near-field phase retrieval algorithm (NfPR)~\cite{boominathan2020phlatcam}. This iterative algorithm alternates between the fields at the mask plane and the sensor plane, enforcing specific constraints at each location: the field amplitude at the mask plane is set to unity, and the intensity of the field at the sensor plane conforms to the target PSF.
The algorithm utilizes forward Fresnel propagation for computing the field from the mask to the sensor plane and applies backward Fresnel propagation to compute from the sensor back to the mask. The inputs required for phase mask optimization include the target PSF, the mask-to-sensor distance (1mm in our setup), and the light wavelength, which in this case is set to 532 nm, a typical wavelength in the mid-visible spectrum.

\begin{figure}[thbp]
\centering
\includegraphics[width=\linewidth]{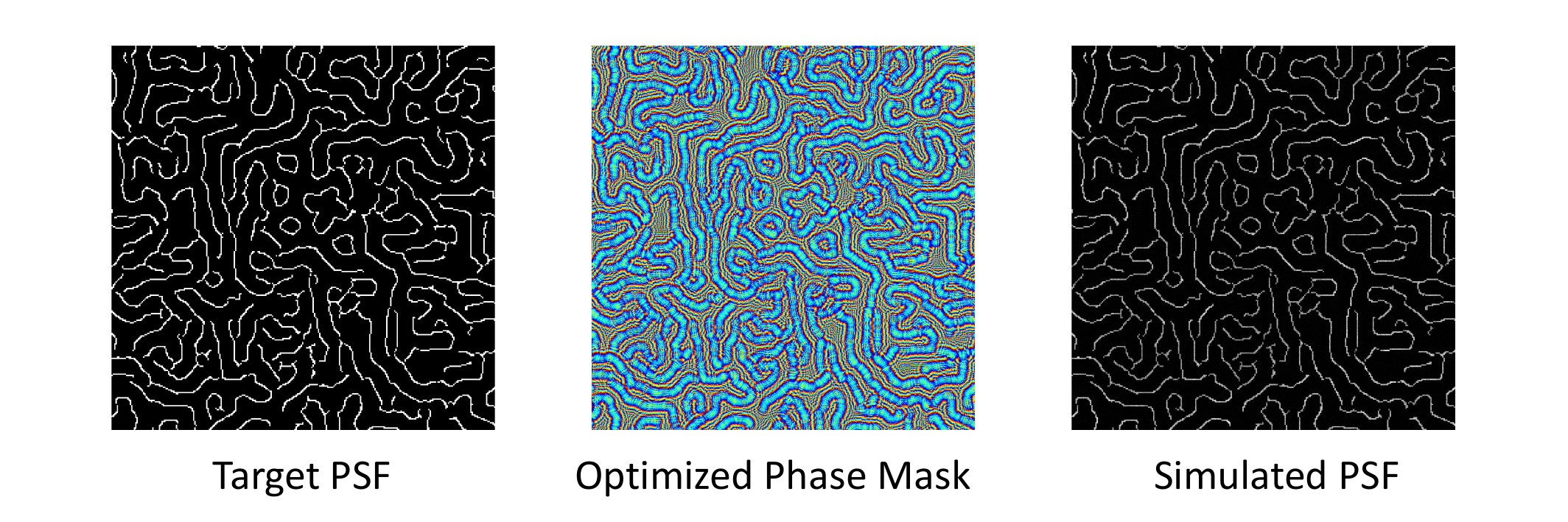}
\caption{
    Visual examples for results of the range space content simulation.
}\label{fig:phlatcam}
\end{figure}

\cref{fig:phlatcam} illustrates the target PSF, the optimized phase mask, and the simulated PSF resulting from the Fresnel propagation method applied to the designed phase mask. Given a desired phase mask, we can accurately simulate the lensless imaging process under various situations.
\section{Range Space Content Approximation}
In the first stage of our method (\cref{sec:svdeconv}), we want to reconstruct the range-space projection $\mathbf{A}^{\dagger}\mathbf{A}\mathbf{x}$ from the lensless measurement $\mathbf{\hat y}$. This reconstruction requires the known target $\mathbf{A}^{\dagger} \mathbf{A} \mathbf{x}$ corresponding to the given image $\mathbf{x}$.
However, for real-world lensless imaging systems, directly obtaining the transfer matrix $\mathbf{A} \in \mathbb{R}^{N^2 \times M^2}$ is computationally infeasible due to its considerable size. Consequently, we resort to simulating the realistic lensless imaging process using an approximate form of $\mathbf{A}$ and emulate its pseudoinverse $\mathbf{A}^{\dagger}$ by inverting this approximation.

In cases where a single point spread function (PSF) $\mathbf{h}$
is calibrated, we approximate the forward imaging process as follows:
\begin{equation}
\mathbf{y}_{\rm sim} = C(\mathbf{h} * \mathbf{x}) + \mathbf{n},
\end{equation}
where $C$ denotes the cropping operation due to the limited sensor size, $*$ represents the convolution operator, and $\mathbf{n}$ represents both sensor and quantization noise. In our practical implementation, the simulated $\mathbf{y}_{\rm sim}$ is quantized to 12 bits, and sensor noise is introduced to achieve a signal-to-noise ratio (SNR) of 30.

The approximation for $\mathbf{A}^{\dagger}$ is derived by reversing the forward process, specifically by applying replicate padding to counteract the cropping and employing a Wiener deconvolution to reverse the convolution:
\begin{equation}
    \mathbf{x}_{\rm range} = D(P(\mathbf{y}_{\rm sim}), \mathbf{h}),
\end{equation}
where the $P$ represent replicate padding and the $D$ represent Wiener deconvolution. 

If multiple calibrated PSFs are available across different fields of view (FoV), the imaging forward process is better approximated by employing a low-rank model to mimic the spatially varying convolution. Correspondingly, the spatially varying deconvolution model discussed in \cref{sec:svdeconv} could be utilized to simulate 
$\mathbf{A}^{\dagger}$.

In our PhlatCam and DiffuserCam setups, we only have a single calibrated PSF. Therefore, we constitute the range-space content of image $\mathbf{x}$ using the simple convolutions. It's important to note that even without multiple calibrated PSFs for a more accurate simulation, the generated range-space content remains a good approximation of the real counterpart. As shown in Section~\ref{sec:analysis}, our simulation still serves as better supervision compared to the original image content in the first stage.

\begin{figure}[thbp]
\centering
\includegraphics[width=\linewidth]{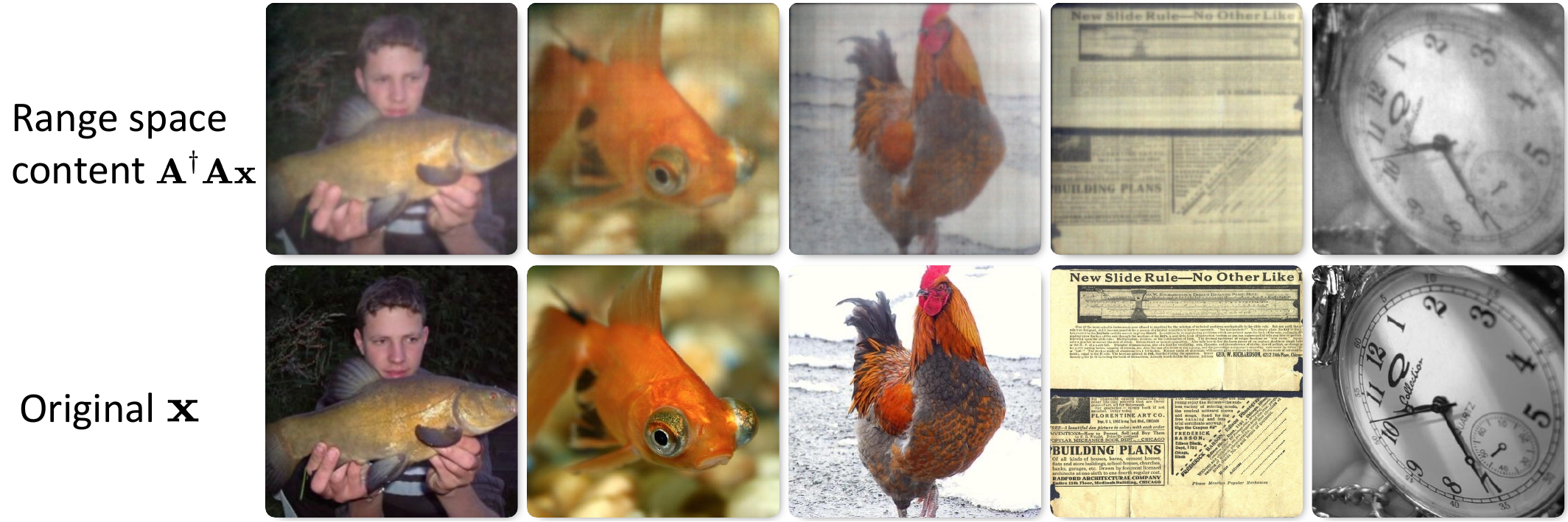}
\caption{
    Visual examples for results of the range space content simulation.
}\label{fig:range_example}
\end{figure}
\begin{figure}[h]
\centering
\includegraphics[width=\linewidth]{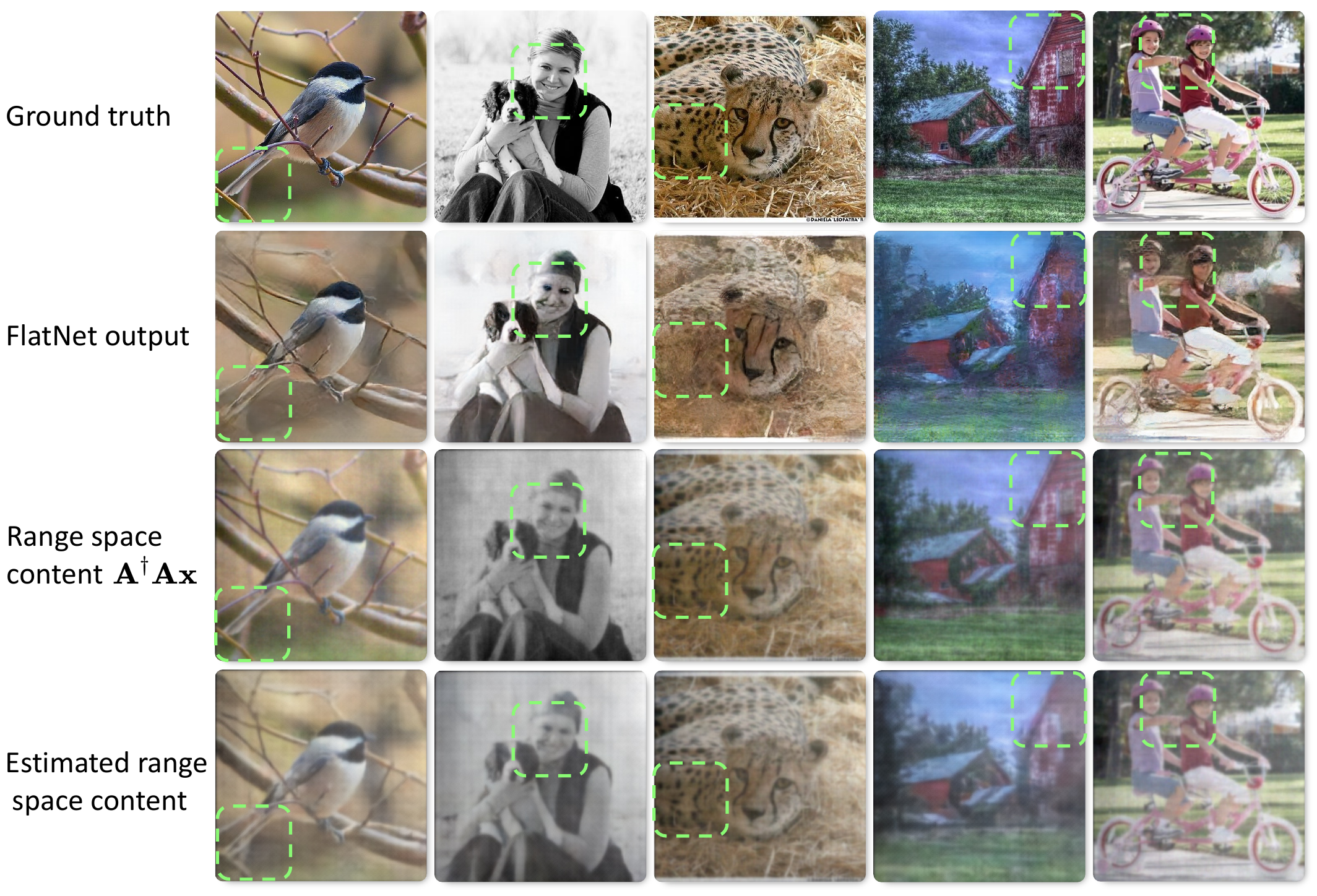}
\caption{
         Examples of range space content estimated by our method. 
}
\label{fig:range_space_result}
\end{figure}
 \cref{fig:range_example} shows the visual examples of the original ground truth $\mathbf{x}$ and the corresponding simulated range space content $\mathbf{A}^{\dagger}\mathbf{A}\mathbf{x}$ of PhlatCam.  It is shown that the range of space content $\mathbf{A}^{\dagger}\mathbf{A}\mathbf{x}$ keeps the low-frequency structure of the original scene and lost some high-frequency details from the lensless imaging process.
 
\cref{fig:range_space_result} shows the visual examples of range space content estimated by our method. The outputs from SVDeconv exhibit high visual consistency with the ground truth, as highlighted in the green boxes, in contrast to those from FlatNet. FlatNet alters the original scene's content and introduces incorrect high-frequency details. Conversely, our method preserves only the low-frequency content in the range space, aligning closely with the original ground truth.

\section{Training Target for the Null-space Diffusion Model}

In \cref{sec:conclusion}, we establish that the goal articulated in  \cref{eq:consistency} corresponds directly to the optimization objective described in  \cref{eq:diffusion} used for diffusion training. We provide a detailed derivation to support this equivalence.

Diffusion models \cite{rombach2022high} can be characterized by a signal-to-noise ratio defined by two sequences \( (\alpha_t)_{t=1}^T \) and \( (\sigma_t)_{t=1}^T \). Given a data sample \( \mathbf{x}_0 \), the forward diffusion process \( q \) can be expressed as:
\begin{equation}
q(\mathbf{x}_t \mid \mathbf{x}_0) = \mathcal{N}(\mathbf{x}_t \mid \alpha_t \mathbf{x}_0, \sigma_t^2 \mathbb{I}),
\end{equation}
where \( \mathcal{N} \) denotes the normal distribution.

The Markov property of the diffusion is given as:
\begin{equation}
q(\mathbf{x}_t \mid \mathbf{x}_s) = \mathcal{N}(\mathbf{x}_t \mid \alpha_{t|s} \mathbf{x}_s, \sigma_{t|s}^2 \mathbb{I}),
\end{equation}
for \( s < t \), with:
\begin{equation}
\alpha_{t|s} = \frac{\alpha_t}{\alpha_s} \quad \text{and} \quad \sigma_{t|s}^2 = \sigma_t^2 - \alpha_{t|s}^2 \sigma_s^2.
\end{equation}

Given \( \mathbf{x}_t = \alpha_t \mathbf{x}_0 + \sigma_t \epsilon \), and a range space condition \( \mathbf{c} = \mathbf{A}^\dagger \mathbf{A} \mathbf{x} \), the diffusion output \( \widetilde{\mathbf{x}} \) can be modeled by \( \widetilde{\mathbf{x}} = x_\theta(\mathbf{x}_t, \mathbf{c}, t) \). To ensure reconstruction consistency such that \( \mathbf{A}(\widetilde{\mathbf{x}} - \mathbf{x}_0) = \mathbf{0} \) in ~\cref{eq:consistency}, the optimization objective is defined as:
\begin{equation}
\min_\theta \mathbb{E}_{\mathbf{x}_0, t, \mathbf{c}, \epsilon \sim \mathcal{N}(0,1)} \left[ \| \mathbf{A}(x_\theta(\mathbf{x}_t, \mathbf{c}, t) - \mathbf{x}_0) \|_2^2 \right].
\end{equation}

Following previous work \cite{ho2020denoising}, we employ the reparameterization of the reconstruction term as a denoising objective:
\begin{equation}
\epsilon_\theta (\mathbf{x}_t,\mathbf{c},t) = \frac{\mathbf{x}_t - \alpha_t x_\theta(\mathbf{x}_t,\mathbf{c},t)}{\sigma_t},
\end{equation}
to minimize:
\begin{equation}
\min_\theta \mathcal{L}_{\rm{null}} = \mathbb{E}_{\mathbf{x}_0, t, \mathbf{c}, \epsilon \sim \mathcal{N}(0,1)} \left[ \| \mathbf{A}(\epsilon_\theta(\mathbf{x}_t, \mathbf{c}, t) - \epsilon) \|_2^2 \right],
\end{equation}
which is exactly the \cref{eq:diffusion}.

Due to the substantial computational cost associated with simulating the transfer matrix $\mathbf{A}$,  an alternative optimization objective can be employed in practice. Specifically, we can approximate the goal described in  \cref{eq:diffusion} using the following optimization formulation:
\begin{equation}
\min_\theta \mathcal{L}_{\rm{null_{app}}} = \mathbb{E}_{\mathbf{x}_0, t, \mathbf{c}, \epsilon \sim \mathcal{N}(0,1)} \left[ \| (\epsilon_\theta(\mathbf{x}_t, \mathbf{c}, t) - \epsilon) \|_2^2 \right].
\end{equation}

\section{Qualitative Results}
\begin{figure}[thbp]
\centering
\includegraphics[width=\linewidth]{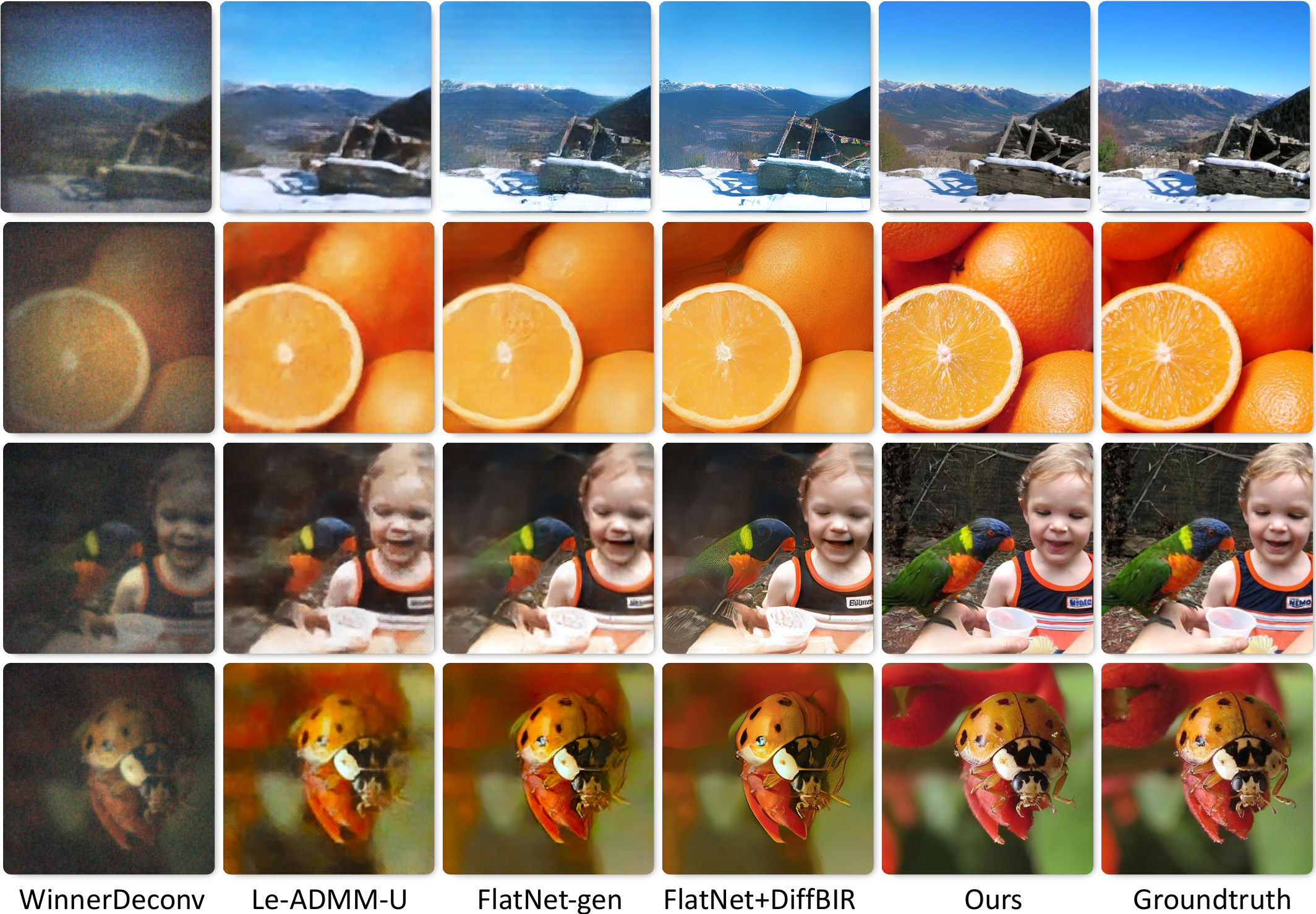}
\caption{
    Qualitative comparison between our method and others on the PhlatCam dataset. 
}
\label{fig:app_result}
\includegraphics[width=\linewidth]{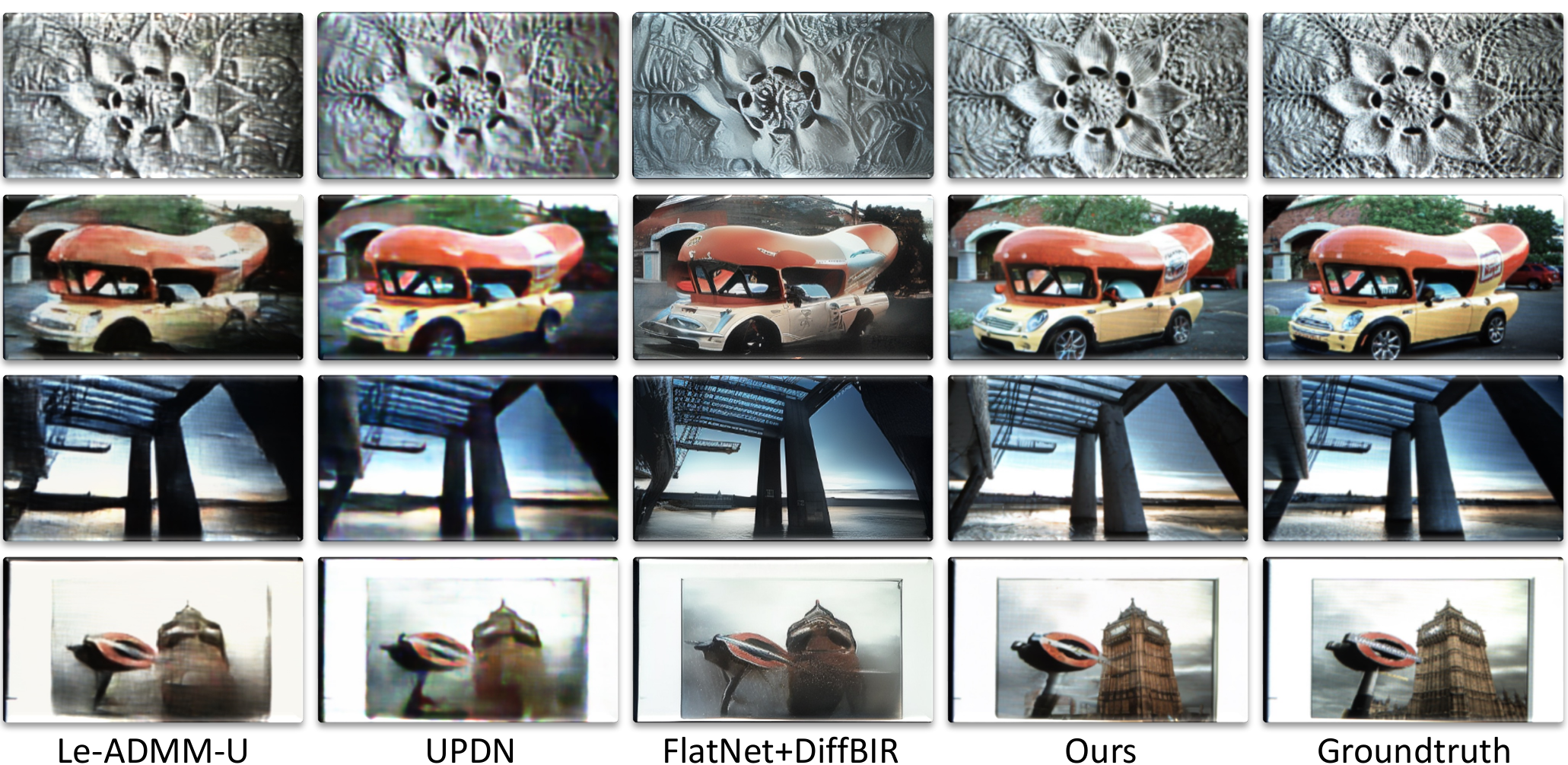}
\caption{
    Qualitative comparison between our method and others on the DiffuserCam dataset. 
\label{fig:app_result2}
}
\vspace{-20pt}

\end{figure}

We present additional qualitative results demonstrating the superior performance of our methods in \cref{fig:app_result} for the PhlatCam dataset and in \cref{fig:app_result2} for the DiffuserCam dataset.

\end{document}